\documentstyle[12pt]{article}
\catcode`@=11
\newcount\@tempcntc
\def\@citex[#1]#2{\if@filesw\immediate\write\@auxout{\string\citation{#2}}\fi
  \@tempcnta\z@\@tempcntb\m@ne\def\@citea{}\@cite{\@for\@citeb:=#2\do
    {\@ifundefined
       {b@\@citeb}{\@citeo\@tempcntb\m@ne\@citea\def\@citea{,}{\bf ?}\@warning
       {Citation `\@citeb' on page \thepage \space undefined}}%
    {\setbox\z@\hbox{\global\@tempcntc0\csname b@\@citeb\endcsname\relax}%
     \ifnum\@tempcntc=\z@ \@citeo\@tempcntb\m@ne
       \@citea\def\@citea{,}\hbox{\csname b@\@citeb\endcsname}%
     \else
      \advance\@tempcntb\@ne
      \ifnum\@tempcntb=\@tempcntc
      \else\advance\@tempcntb\m@ne\@citeo
      \@tempcnta\@tempcntc\@tempcntb\@tempcntc\fi\fi}}\@citeo}{#1}}
\def\@citeo{\ifnum\@tempcnta>\@tempcntb\else\@citea\def\@citea{,}%
  \ifnum\@tempcnta=\@tempcntb\the\@tempcnta\else
   {\advance\@tempcnta\@ne\ifnum\@tempcnta=\@tempcntb \else \def\@citea{--}\fi
    \advance\@tempcnta\m@ne\the\@tempcnta\@citea\the\@tempcntb}\fi\fi}
\catcode`@=12

\def\barr{\begin{array}}
\def\earr{\end{array}}
\def\beq{\begin{equation}}
\def\eeq{\end{equation}}
\def\bea{\begin{eqnarray}}
\def\eea{\end{eqnarray}}
\def\bmath{\begin{displaymath}}
\def\emath{\end{displaymath}}
\def\bq{\begin{quote}}
\def\eq{\end{quote}}

\def\cA{{\cal A}}

\def\cP{{\cal P}}
\def\cT{{\cal T}}

\def\lh{\lambda_h}
\def\lR{\lambda_R}
\def\PL{\mbox{P}_L}
\def\PR{\mbox{P}_R}
\def\veps{\varepsilon}
\def\apprle{\hspace{-0.1cm}\stackrel{\displaystyle <}{\sim}}

\def\slash#1{\setbox0=\hbox{$#1$}#1\hskip-\wd0\hbox to\wd0{\hss\sl/\/\hss}}
\def\lNN{\lambda_N}
\def\lNi{\lambda_{N_i}}
\def\lNj{\lambda_{N_j}}
\def\lZ{\lambda_Z}
\def\lN1{\lambda_{N_1}}

\def\npb#1{{\em Nucl.\ Phys.\ }{\bf B#1}}
\def\plb#1{{\em Phys.\ Lett.\ }{\bf A#1}}
\def\plb#1{{\em Phys.\ Lett.\ }{\bf B#1}}
\def\prl#1{{\em Phys.\ Rev.\ Lett.\ }{\bf #1}}

\def\prd#1{{\em Phys.\ Rev.\ }{\bf D#1}}

\def\ptp#1{{\em Prog.\ Theor.\ Phys.\ }{\bf #1}}

\def\zpc#1{{\em Z.\ Phys.\ }{\bf C#1}}
\begin{document}

\begin{flushright}
RAL/94-089\\
FTUV/94-56
\end{flushright}

\begin{center}
{\Large{\bf Constraints from Lepton Universality at the {\em Z} Peak}}\\[0.3cm]
{\Large{\bf on Unified Theories}}\\[1.5cm]
{\large J.~Bernab\'eu}$^a${\large ~and~A.~Pilaftsis}$^b$\footnote[1]{
E-mail address: pilaftsis@v2.rl.ac.uk}\\[0.4cm]
{\em $^a$Departament de Fisica Te\'orica, Univ.~de Val\'encia, and
IFIC,}\\[-0.1cm]
{\em Univ.~de Val\'encia--CSIC, E-46100 Burjassot (Val\'encia), Spain}\\[0.2cm]
{\em $^b$Rutherford Appleton Laboratory, Chilton, Didcot, Oxon, OX11 0QX,
UK}
\end{center}
\vskip1.5cm
\centerline {\bf ABSTRACT}
We suggest the use of a universality-breaking observable based
on lepton asymmetries as derived from the left-right asymmetry and the
$\tau$ polarization at the $Z$ peak, which can efficiently
constrain the parameter space of unified theories. The new observable
is complementary to the leptonic partial width differences and it
depends critically on the chirality of a possible non-universal $Z$-boson
coupling to like-flavour leptons. The LEP/SLC potential of probing
universality violation is discussed in representative low-energy
extensions of the Standard Model (SM) that could be derived by
supersymmetric grand unified theories, such as the SM with
left-handed and/or right-handed neutral isosinglets, the left-right
symmetric model, and the minimal supersymmetric SM.

\newpage

Supersymmetric (SUSY) grand unified theories (GUTs), such as the
SUSY-$SU(5)$ model~\cite{SUSYGUT}, have received much attention due to the
recent observation that the prediction obtained for the electroweak
mixing angle, $\sin^2\theta_w (\equiv s^2_w)$, is in
excellent agreement with its value measured experimentally.
Another interesting feature is that the supersymmetric nature of a
SUSY-GUT model has the tendency to drive the unification point to
higher values than usual GUTs by one or two orders of magnitude,
which prevents proton from decaying too rapidly.
The low-energy limit of a SUSY-GUT scenario depends crucially on the
possible representations of the chiral multiplets contained and the
details of the breaking mechanism from the unification scale down to
the electroweak one.
For instance, the SM with right-handed neutrinos could be viewed as a
conceivable low-energy realization of certain SUSY-GUTs,
{\em e.g.}~the SUSY versions of the models discussed in Ref.~\cite{YAN}.
Such a minimal extension of the SM allows the presence of high Dirac
mass terms without contradicting constraints on the
light neutrino masses~\cite{ZPC,BW}.
As an immediate consequence, the one-loop vertex
function relevant for the lepton-flavour-violating decays of the
Higgs~\cite{APetal} and~$Z$ bosons~\cite{KPS} shows a strong quadratic
dependence on the heavy neutrino mass, leading to rates that could be
probed at the CERN Large Electron Positron Collider (LEP).
For this purpose, an observable $U_{br}$ measuring deviations from lepton
universality has been suggested in Ref.~\cite{BKPS} in order to
effectively constrain possible nondecoupling effects originating from
heavy Majorana neutrinos.

In this note we introduce a universality-breaking observable
based on lepton asymmetries at the $Z$ peak and discuss its
phenomenological implications within the framework of three
representative extensions of the SM that could naturally be
derived by SUSY-GUTs:
(i) the SM with left-handed and/or right-handed neutral isosinglets,
(ii) the left-right symmetric model, and (iii) the minimal SUSY-SM.
The new observable is complementary to $U_{br}$ and depends explicitly on
the chirality of a possible non-universal $Zl\bar{l}$ coupling where $l$
denotes the charged leptons $e$, $\mu$, and $\tau$.

In the limit $m_l\to 0$, the transition amplitude of the decay
$Z\to l\bar{l}$ can generally be given by
\beq
\cT_l\ =\ \frac{ig_w}{2c_w}\, \veps^\mu_Z\, \bar{u}_l\gamma_\mu [g^l_L\PL\
+\ g^l_R\PR ] v_l,
\eeq 
where $g_w$ is the usual electroweak coupling constant,
$\PL(\PR)=(1-(+)\gamma_5)/2$,
$c^2_w=1-s^2_w=M^2_W/M^2_Z$,
and $\veps^\mu_Z$ and $u_l$ ($v_l$) are the polarization vector of
the $Z$ boson and the Dirac spinor of $l$ ($\bar{l}$), respectively.
Furthermore, we have defined $g^l_{L,R}=g_{L,R}+\delta g^l_{L,R}$, where
$g_L=1-2s^2_w$ and $g_R=-2s^2_w$ are the values at tree level, and
$\delta g^l_{L,R}$ are obtained beyond the Born approximation.
To the first nonvanishing order of perturbation theory, the above
parametrization enables us to express the universality-breaking
parameter $U_{br}^{(ll')}$~\cite{BKPS} as follows:
\beq
U_{br}^{(ll')}\ =\ \frac{\Gamma (Z\to l\bar{l})\ -\ \Gamma (Z\to l'\bar{l}')}
{\Gamma (Z\to l\bar{l})\ +\ \Gamma (Z\to l'\bar{l}')}\
= \  \frac{g_L (\delta g_L^l - \delta g^{l'}_L)\ +\ g_R (\delta g^l_R
- \delta g_R^{l'} )}{g_L^2\ +\ g^2_R}.
\eeq 
In Eq.~(2),  the known phase-space
corrections coming from the masses of the charged leptons $l$ and $l'$
have been  subtracted. To make contact with the
corresponding observable given in~\cite{PDG}, one can easily derive the
relation: $\Gamma (Z\to l\bar{l})/\Gamma (Z\to l'\bar{l}')
= 2 U^{(ll')}_{br} + 1$. On the other hand, in the massless limit of
final leptons, the lepton asymmetry $\cA_l$ is given by
\beq
\cA_l \ =\ \frac{\Gamma (Z\to l_L \bar{l})\ -\ \Gamma (Z\to l_R \bar{l})}{
\Gamma (Z\to l \bar{l})}\
=\ \frac{ g_L^2 -g^2_R + 2(g_L\delta g^l_L - g_R\delta g^l_R) }{
g^2_L + g^2_R + 2(g_L\delta g^l_L + g_R\delta g^l_R)}.
\eeq 
Note that $\cA_e$ measured at LEP should equal the left-right asymmetry,
   $\cA_{LR}$, obtained
at the Stanford Linear Collider (SLC).
In view of the recent discrepancy of
about $2\sigma$ between SLC and LEP results~\cite{SLD}
for $\cA_e$ and $\cA_\tau$, respectively, we suggest using
a universality-breaking parameter involving lepton asymmetries
\beq
\Delta\cA_{ll'}\ =\ \frac{\cA_l\ -\ \cA_{l'}}{\cA_l\ +\ \cA_{l'}}\
=\ \frac{1}{\cA_l^{(SM)}} \Big( U_{br}^{(ll')}(\mbox{L})\ -\
U_{br}^{(ll')}(\mbox{R})
\Big)\ -\ U_{br}^{(ll')}
\eeq 
where $\cA_l^{(SM)}$ is the SM  lepton asymmetry and
$U_{br}^{(ll')}=U_{br}^{(ll')}(\mbox{L})+U_{br}^{(ll')}(\mbox{R})$, with
$U_{br}^{(ll')}(\mbox{L})
=g_L(\delta g_L^l- \delta g_L^{l'})/(g^2_L+g^2_R)$
and $U_{br}^{(ll')}(\mbox{R})
=g_R(\delta g_R^l- \delta g_R^{l'})/(g^2_L+g^2_R)$.
It should be stressed that requiring $U_{br}^{(ll')}=0$ does
{\em not necessarily} imply $\Delta\cA_{ll'}=0$. Moreover,
contrary to $\cA_{LR}$ considered in Ref.~\cite{PT}, our observable
$\Delta\cA_{ll'}$ does not depend explicitly on universal electroweak
oblique parameters. In the following,
we will calculate $\Delta\cA_{ll'}$ [or equivalently
$U_{br}^{(ll')}(\mbox{L})$ and $U_{br}^{(ll')}(\mbox{R})$] in
three illustrative and minimal extensions of the SM as mentioned
above.

\noindent
{\em (i) The SM with left-handed and/or right-handed neutral
isosinglets}. Such a model can be obtained by adding right-handed
neutrinos to the field content of the SM.
We adopt the notation of Ref.~\cite{ZPC} for the charged-
and neutral-current interactions. The coupling of the $W$ boson to a
charged lepton $l$ and heavy Majorana neutrinos $N_i$ is proportional
to the mixing $B_{lN_i}$. The corresponding $ZN_iN_j$ coupling is
governed by the mixing matrix $C_{N_iN_j}$. For a model with two-right
handed neutrinos, for example, we have
\beq
B_{lN_1}\ =\ \frac{\rho^{1/4} s^{\nu_l}_L}{\sqrt{1+\rho^{1/2}}}\ , \qquad
B_{lN_2}\ =\ \frac{i s^{\nu_l}_L}{\sqrt{1+\rho^{1/2}}}\ ,
\eeq 
where $\rho=m^2_{N_2}/m^2_{N_1}$ is the square of the
mass ratio between the two heavy
Majorana neutrinos $N_1$ and $N_2$ predicted in such a model,
and $s^{\nu_l}_L$ is defined as~\cite{LL}:
$(s^{\nu_l}_L)^2\equiv \sum\limits_{j=1}^{n_R} |B_{lN_j}|^2$.
Furthermore, the mixings $C_{N_iN_j}$ can be obtained by
$\sum\limits_{l=1}^{n_G} B_{lN_i}^{\ast}B_{lN_j}  =  C_{N_iN_j}$.
The mixing angles $(s^{\nu_i}_L)^2$ are directly constrained by
low-energy and LEP data~\cite{Roulet,BGKL}.
Although some of the constraints could be model-dependent,
in our analysis we will use the conservative upper limits~\cite{BGKL}:
$(s^{\nu_e}_L)^2$, $(s^{\nu_\mu}_L)^2 < 0.01$, and
$(s^{\nu_\tau}_L)^2 < 0.06$.
Another limitation to the parameters of our model comes from the
requirement of the validity of perturbative unitarity that can be violated
in the limit of large heavy-neutrino masses.
A qualitative estimate for the latter may be obtained by requiring that
the total widths, $\Gamma_{N_i}$, and masses of neutrino fields $N_i$
satisfy the inequality $\Gamma_{N_i}/m_{N_i}< 1/2$~\cite{BKPS}.
Taking into account dominant and subdominant nondecoupling
contributions when $\lN1 = m^2_{N_1}/M^2_W \gg 1$ and
$\rho=m^2_{N_2}/m^2_{N_1} \ge 1$~\cite{BKPS},
we find that
\bea
U_{br}^{(ll')}\ =\ U_{br}^{(ll')}(\mbox{L}) &=& -\frac{\alpha_w}{8\pi}\,
\frac{g_L}{g^2_L+g^2_R}\, \Big( (s_L^{\nu_l})^2-(s_L^{\nu_{l'}})^2 \Big)
\Bigg[ 3\ln\lN1
               \nonumber\\
   &+&\sum_{i=1}^{n_G}\ (s_L^{\nu_i})^2\;
                  \frac{\lN1 }{(1+\rho^{\frac{1}{2}})^2}\Bigg( 3\rho
                   +\frac{\rho-4\rho^{\frac{3}{2}}+\rho^2}
                         {2(1-\rho)}\ln\rho\Bigg)\Bigg].
\eea 
As expected, in $SU(2)_L\otimes U(1)_Y$ models,
the nature of a possible universality breaking is pure left-handed.

Another attractive low-energy scenario is an extension of the SM
inspired by certain GUTs~\cite{WW} and superstring theories~\cite{EW,BSVMV},
in which left-handed neutral singlets in addition to the right-handed
neutrinos are present.
In this scenario, the light neutrinos are strictly massless
to all orders of perturbation theory~\cite{WW}, when $\Delta L=2$
operators are absent from the Yukawa sector.
The minimal case with one left-handed and one right-handed chiral singlets
can effectively be recovered by the SM with two right-handed
neutrinos when taking the degenerate mass limit for the two heavy Majorana
neutrinos in Eq.~(6). In this way, we obtain
\beq
U_{br}^{(ll')}(\mbox{L}) = -\frac{\alpha_w}{8\pi}\,
\frac{g_L}{g^2_L+g^2_R}\, \Big( (s_L^{\nu_l})^2-(s_L^{\nu_{l'}})^2 \Big)
\Big[ 3\ln\lNN\ +\ \sum_{i=1}^{n_G}\ (s_L^{\nu_i})^2\; \lNN \Big].
\eeq 
In Table 1, we present numerical results for both scenarios discussed
above by assuming $m_{N_1}\simeq m_{N_2} = m_N$, in which case Eq.~(6)
leads to Eq.~(7). The present experimental
upper bound on $U_{br}^{(ll')}$ is $|U_{br}^{(ll')}| < 5.\ 10^{-3}$~\cite{PDG},
which automatically sets an upper limit on $|\Delta\cA_{ll'}| \apprle 3\%$
since $U_{br}^{ll'}(\mbox{R})=0$. The experimental bound on
$\Delta\cA_{ll'}$ depends on the value of $\cA_l^{(SM)}$ we use.
In fact, it is $\cA_{LR}=0.1637\pm 0.0076$~\cite{SLD}
and $\cA_\tau(\cP_\tau ) = 0.143\pm 0.010$ ($\cA_e (\cP_\tau )
= 0.135 \pm 0.011$)~\cite{CERN} from measurements at SLC and
LEP, respectively. The present experimental non-vanishing value of
$\Delta\cA_{\tau e}$ has become the subject of some recent theoretical
works~\cite{Ross}. Our aim is to show the sensitivity of $\Delta\cA_{\tau e}$
to nonoblique flavour-dependent diagonal interactions of the $Z$ boson
induced by new physics.
Table~1 shows theoretical predictions for
$\Delta\cA_{ll'}$ close to  values of phenomenological interest.

\noindent
{\em (ii) The left-right symmetric model}. This model extends
the gauge sector of the SM by an extra isospin $SU(2)_R$ group.
For simplicity, we have worked out the realistic
case (d) in~\cite{GGMKO}, in which the vacuum expectation values of
the left-handed Higgs triplet $\Delta_L$ and that of $\phi^0_2$
in the Higgs bi-doublet vanish. The model can give rise to both a
left-handed and a right-handed non-universal $Zl\bar{l}$ coupling.
The expression for $U_{br}^{(ll')}(\mbox{L})$ coincides with the
one given in Eq.~(6), while the dominant nondecoupling contributions
to $U_{br}^{(ll')}(\mbox{R})$ can be obtained by calculating
the Feynman graphs shown in Fig.~1. In the applicable limit
where the charged gauge bosons $W^\pm_R$ associated
with the group $SU(2)_R$ and the charged Higgs bosons $h^\pm$ are
much heavier than the $Z$ boson, we find
\bea
U_{br}^{(ll')}(\mbox{R}) &=& \frac{\alpha_w}{8\pi}\,
\frac{g_R}{g^2_L+g^2_R}\, \Big( B^R_{lN_i}B^{R\ast}_{lN_j}-
B^R_{l'N_i}B^{R\ast}_{l'N_j} \Big) \sqrt{\lNi\lNj}\nonumber\\
&&\times\ \Big[\, \delta_{ij} F_1\ +\ C^L_{N_iN_j}F_2\ +\
C^{L\ast}_{N_iN_j}F_3\, \Big],
\eea 
where $F_1$, $F_2$, and $F_3$ are form factors given by
\bea
F_1 &=& 4s^2_\beta [ I(\lR,\lR,\lNi)\ -\ I(\lR,\lh,\lNi) ], \\
F_2 &=& 2[ I(\lR,\lh,\lNi)\ +\ I(\lR,\lh,\lNj )\ -\
I(\lNi,\lNj,\lR)]\nonumber\\
&&+\ s^2_\beta [ L(\lh,\lh,\lNi )\ +\ L(\lh,\lh,\lNj )\ -\ L(\lR,\lh,\lNi )\
-\ L(\lR,\lh,\lNj)\nonumber\\
&& +\ L(\lNi,\lNj,\lR)\ -\ L(\lNi,\lNj,\lh ) ]\ +\ \frac{s^2_\beta}{c^2_\beta}
\, [ L(\lNi,\lNj,\lh )\ -\ L( 0, \lNj,\lh )\nonumber\\
&& -\ L(\lNi,0,\lh )\ +\ L(0,0,\lh ) ], \\
F_3 &=& -\, \frac{2}{\sqrt{\lNi\lNj}}\, [ L(\lNi,\lNj,\lR )\ -\
L(0,\lNj,\lR )\ -\ L( \lNi,0,\lR )\ + L(0,0,\lR )] \nonumber\\
&&+\ s^2_\beta\, \sqrt{\lNi\lNj}\, I(\lNi,\lNj,\lR ),
\eea 
with $\lR=1/s^2_\beta=M^2_R/M^2_W$ and $\lh = M^2_h/M^2_W$.
In Eq.~(8), $B^R$ and $C^L$ (no left-right mixing is assumed)
are mixing matrices parametrizing the couplings $W_RlN$ and $ZNN$,
respectively.
The loop functions $I$ and $L$ in Eqs.~(9)--(11) may conveniently
be defined as
\bea
I(a,b,c) &=& \int_0^1\int_0^1 y\, dxdy/\{ [a(1-x)+bx]y+c(1-y)\},\nonumber\\
L(a,b,c) &=& \int_0^1\int_0^1 y\, dxdy\, \ln\{ [a(1-x)+bx]y+c(1-y)\}.\nonumber
\eea
The value of $U_{br}$(R) depends on many kinematic variables, {\em i.e.},
the masses of heavy neutrinos (in our numerical estimates we fix them to
be $m_N=4$~TeV), the $W_R$-boson mass ($M_R$), and the charged Higgs
mass ($M_h$). Typically, we have set $(s^{\nu_\tau}_L)^2 =0.05$
and $(s^{\nu_e}_L)^2 = 0.01$.
As can be seen from Table 1, despite the fact that $U_{br}^{(\tau e)}$ could
be unobservably small of the order of $10^{-3}$ in this model,
$\Delta\cA_{\tau e}$ can be as large as $10\%$ well within the experimental
reach of LEP and SLC.

\noindent
{\em (iii) The minimal supersymmetric SM}. In this minimal scenario,
nonvanishing values for $U_{br}^{(ll')}(\mbox{L})$ and
$U_{br}^{(ll')}(\mbox{R})$ can be induced by left-handed and
right-handed scalar leptons (denoted as $\tilde{l}_L$, $\tilde{l}_R$)
as well as scalar neutrinos. To generate a non-zero non-universal
$Zl\bar{l}$ coupling, it is sufficient that two left-handed or
right-handed scalar leptons, say $\tilde{l}$ and $\tilde{l}'$,
are not degenerate.
In addition, we will consider the SUSY limit
of the gaugino sector, where only explicit SUSY-breaking
scalar-lepton mass terms are present. Only two neutralinos, the
photino $\tilde{\gamma}$ and the "ziggsino"
$\tilde{\zeta}$ with mass $m_{\tilde{\zeta}}=M_Z$, will then
contribute as shown in Fig.~2. "Ziggsino" is a Dirac fermion composed
from degenerate Majorana states of a zino $\tilde{z}$ (the SUSY partner
of the $Z$ boson) and one of
the higgsino fields. For the sake of illustration, we will further
assume that only one scalar lepton $\tilde{l}$ is relatively light
whereas the others,
{\em e.g.}~$\tilde{l}'$, are much heavier than $M_Z$. Since
the decoupling theorem in softly broken SUSY theories will be
operative~\cite{LMS}, we neglect quantum effects of
$\tilde{l}'$. A straightforward calculation then gives
\bea
U_{br}^{(ll')}(\mbox{L}) &=& -\; \frac{\alpha_w}{8\pi}\,
\frac{g_L^4\cos 2\theta_L}{g^2_L+g^2_R}\ \Bigg[ \frac{g^2_R}{g^2_L}
\int_0^1\int_0^1 dxdy\; y\; \ln\left( 1-
\frac{\lZ}{\lambda_{\tilde{l}_L}}yx(1-x)\right)\nonumber\\
&&+\ \lZ\int_0^1\int_0^1 dx dy\; y\; \ln\left(
\frac{\lambda_{\tilde{l}_L}y + \lZ [1-y-y^2x(1-x)]}{\lambda_{\tilde{l}_L}y
+ \lZ (1-y)} \right)\Bigg], \\
U_{br}^{(ll')}(\mbox{R}) &=& -\; \frac{\alpha_w}{8\pi}\,
\frac{g_R^4\cos 2\theta_R}{g^2_L+g^2_R}\ \Bigg[
\int_0^1\int_0^1 dxdy\; y\; \ln\left( 1-
\frac{\lZ}{\lambda_{\tilde{l}_R}}yx(1-x)\right)\nonumber\\
&&+\ \lZ\int_0^1\int_0^1 dx dy\; y\; \ln\left(
\frac{\lambda_{\tilde{l}_R}y + \lZ [1-y-y^2x(1-x)]}{\lambda_{\tilde{l}_R}y
+ \lZ (1-y)} \right)\Bigg].
\eea 
Here, $\lZ=M^2_Z/M^2_W$, $\lambda_{\tilde{l}_L}=m^2_{\tilde{l}_L}/M^2_W$,
$\lambda_{\tilde{l}_R}=m^2_{\tilde{l}_R}/M^2_W$, and $\theta_L$
($\theta_R$) is a mixing angle between the two left-handed
(right-handed) scalar leptons $\tilde{l}_L$ ($\tilde{l}_R$) and
$\tilde{l}'_L$ ($\tilde{l}'_R$). As has been displayed in Table 1, numerical
estimates reveal that the universality-violating observables
$U_{br}$ and $\Delta\cA$ are predicted to be
no much bigger than $10^{-3}$. However, in SUSY-GUTs, numerous
new fields and the corresponding SUSY partners are in general present,
yielding group theoretical factors and quantum effects larger
than the minimal SUSY-SM. For example, in SUSY models with right-handed
neutrinos, one may get enhancements coming from the SUSY Yukawa sector,
when the charged higgsinos couple to leptons and scalar neutrinos.
One may therefore expect that
$\Delta\cA_{\tau e}$ could reach an experimentally accessible
level $\sim 10^{-2}$.

In conclusion, we have considered  the observable $\Delta\cA_{ll'}$
in Eq.~(4) based on lepton asymmetries, which can effectively show
a discrepancy between the left-right asymmetry at SLC and $\tau$
polarization at LEP originating from
new physics. $\Delta\cA_{ll'}$ is sensitive to the nature of
chirality of a possible nonuniversal
$Zl\bar{l}$ coupling and is hence complementary to $U_{br}$ discussed
in~\cite{BKPS}. To precisely demonstrate this, we have analyzed
conceivable low-energy scenarios of unified theories, such as
the SM with neutral isosinglets, the
left-right symmetric model and the minimal SUSY model, which can induce
sizeable
values for $\Delta\cA_{\tau e}$ at the experimental visible level of
$5-10\%$, whereas non-SM signals through the usual observable
$U_{br}$ may turn out to be rather small. As is seen from Table~1,
the sign of $\Delta\cA_{\tau e}$ will provide a discrimination among
the various theoretical scenarios beyond the SM.


\newpage

\centerline{\Large\bf Figure and Table Captions }

\newcounter{fig}
\begin{list}{\bf\rm Fig. \arabic{fig}: }{\usecounter{fig}
\labelwidth1.6cm \leftmargin2.5cm \labelsep0.4cm \itemsep0ex plus0.2ex }

\item Feynman diagrams generating a right-handed non-universal $Zl\bar{l}$
coupling in the left-right symmetric model. Contributions from wave-function
renormalization constants of the charged leptons have been considered but not
displayed.

\item Feynman graphs contributing to a left-handed and right-handed\\
non-universal $Zl\bar{l}$ coupling. In addition, selfenergies of
external leptons $l$ have been taken into account.

\end{list}
\newcounter{tab}
\begin{list}{\bf\rm \hspace{-0.1cm}Tab. \arabic{tab}: }{\usecounter{tab}
\labelwidth1.6cm \leftmargin2.5cm \labelsep0.4cm \itemsep0ex plus0.2ex }

\item  Numerical estimates of the universality-breaking observables
$U^{\tau e}_{br}$(L), $U^{\tau e}_{br}$(R),\\ and $\Delta\cA_{\tau e}$
in the context of models (i), (ii), and (iii) discussed in the text.
We have used the value $s^2_w=0.232$.
\end{list}

\newpage

\centerline{\Large\bf Table 1}
\vspace{0.5cm}
\begin{tabular*}{11.8823cm}{|rr||r|r||l|}
\hline
 Gauge&Models\hspace{0.5cm}& $U^{\tau e}_{br}(\mbox{L})\ $ &
$U_{br}^{\tau e}(\mbox{R})\  $ &
$\Delta\cA_{\tau e}$ \\
\hline\hline
(i) ~~$(s^{\nu_\tau}_L)^2$ & $m_N\ [$TeV$]$ & &  & \\
      0.060 & 4.0 & $-1.5\ 10^{-2}$ & 0 &$-9.1\ 10^{-2}$\\
      0.035 & 4.0 & $-4.0\ 10^{-3}$ & 0 &$-2.4\ 10^{-2}$ \\
      0.020 & 4.0 & $-2.0\ 10^{-3}$ & 0 &$-1.2\ 10^{-2}$ \\
\hline
(ii) $M_R\ [$TeV$]$ & $M_h\ [$TeV$]$ & & & \\
0.4 & 5    & $-1.\ 10^{-2}$&$7.7\ 10^{-3}$&$-0.13$ \\
0.4 & 25   &   "~~~"~~~~   &$1.0\ 10^{-2}$&$-0.14$ \\
0.4 & 50   &   "~~~"~~~~   &$1.5\ 10^{-2}$&$-0.18$ \\
1.0 & 5    &   "~~~"~~~~   &$1.2\ 10^{-3}$&$-0.16$ \\
1.0 & 100  &   "~~~"~~~~   &$3.2\ 10^{-3}$&$-0.10$ \\
1.0 & 1000 &   "~~~"~~~~   &$6.0\ 10^{-3}$&$-0.12$ \\
\hline
(iii) ~$\theta_L=0$ & $m_{\tilde{l}}\ [$GeV$]$ & & & \\
$\theta_R=\frac{\pi}{2}$ &
              45    & $1.1\ 10^{-4}$ & $-6.4\ 10^{-5}$ &
                                       $1.2\ 10^{-3}$ \\
$m_{\tilde{l}_L}=m_{\tilde{l}_R}$&
              60    & $5.5\ 10^{-5}$ & $-3.1\ 10^{-5}$ &
                                       $6.1\ 10^{-4}$ \\
$=m_{\tilde{l}}$    & 100    & $2.0\ 10^{-5}$ & $-1.1\ 10^{-5}$ &
                                       $2.2\ 10^{-4}$ \\
\hline
\end{tabular*}

\end{document}